\documentclass[twocolumn,showpacs,superscriptaddress]{revtex4}

\usepackage{amsmath}
\usepackage{bm}
\usepackage{array}
\usepackage{graphicx}
\usepackage{dcolumn}   % Align table columns on decimal point
\usepackage{amssymb}
\usepackage[english]{babel}

\begin{document}

%\preprint{PRL}
\preprint{PRA}

\title{Measurable quantum geometric phase from a rotating single spin}
\date{\today}

\author{D. Maclaurin}
\affiliation{School of Physics, The University of Melbourne, Parkville,
3010, Australia}
\affiliation{Department of Physics, Harvard University, Cambridge, MA 01238, USA}
\author{M.W. Doherty}
\affiliation{School of Physics, The University of Melbourne, Parkville,
3010, Australia}
\affiliation{Centre for Quantum Computation and Communication Technology, School of Physics, The University of Melbourne, Parkville 3010, Australia}
\author{L.C.L. Hollenberg}
\affiliation{School of Physics, The University of Melbourne, Parkville,
3010, Australia}
\affiliation{Centre for Quantum Computation and Communication Technology, School of Physics, The University of Melbourne, Parkville 3010, Australia}
\author{A.M. Martin}
\affiliation{School of Physics, The University of Melbourne, Parkville,
3010, Australia}

\begin{abstract}
We demonstrate that the internal magnetic states of a single nitrogen-vacancy defect, within a rotating  diamond crystal, acquire geometric phases. The geometric phase shift is manifest as a relative phase between components of a superposition of magnetic substates. We demonstrate that under reasonable experimental conditions  a phase shift of up to four radians could be measured. Such a measurement of the accumulation of a geometric phase, due to macroscopic rotation, would be the first for a single atom-scale quantum system.   
%We propose an experiment that would produce and measure a large geometric phase in  a solid state system under macroscopic rotation. Specifically, a diamond crystal is mounted on a spinning spindle. The internal magnetic states of a single nitrogen-vacancy defect, within the diamond crystal, acquire geometric phases due to the macroscopic rotation. The geometric phase shift is manifest as a relative phase, of upto four radians, between components of a superposition of magnetic substates. To our knowledge such a measurement of the accumulation of a geometric phase, due to macroscopic rotation, would be the first for a single atom-scale quantum system.   
\end{abstract}
\maketitle

%\section{Introduction}
The quantum geometric phase is at the core of our understanding of the non-intuitive quantum view of the world, with historical origins going back to the original paper by Aharonov and Bohm \cite{Aharonov59}, in 1959. However, while there have been many experimental demonstrations on ensembles of quantum systems, no measurement of a quantum phase on an individual quantum system, undergoing macroscopic rotation, has been performed.  In this paper we analyze the appearance of the geometric phase in defects in diamond and show how the measurement of the geometric phase of a single quantum spin, undergoing macroscopic rotation, is now in reach. 

%In addition to the fundamental interest of directly measuring a geometric phase shift in a single quantum state, recent work on the measurement of nano-diamond rotation in live human tumour cells \cite{Mcquinness11} highlights the need to quantify geometric phase shift effects in the use of nano-diamonds as high precision translational and rotational sensors.  
  
%In 1959 Aharonov and Bohm published a paper \cite{Aharonov59} in which they demonstrated a surprisingimplication of quantum mechanics: they showed that electromagnetic potentials can
%affect the dynamics of a particle which moves in a field-free region of space. An electron
%travelling inside a charged cylinder acquires a phase due to the cylinderÕs voltage, though
%the electric field inside the cylinder is zero. Similarly, an electron travelling near a solenoid
%acquires a phase dependent on the vector potential encircling the solenoid, even though
%the magnetic field outside the solenoid vanishes.

%These findings initially received surprise, even skepticism. They suggested either nonlocality
%in quantum mechanics or otherwise a physical basis for electromagnetic potentials,
%which had previously been treated as mathematical conveniences only. Experiments soon
%verified the predictions \cite{Chamber60} and the Aharonov-Bohm effect became a textbook example
%of the counter-intuitiveness of quantum mechanics.

In 1984, M.V. Berry developed an elegant and powerful mathematical framework
which established the Aharonov-Bohm effect as just one instance of a far more general class of
phenomena \cite{Berry84}. Berry considered the evolution of a system under a Hamiltonian which
is adiabatically changed over time. He showed that the state of such a system
acquires a phase which is geometrical in nature. The phase depends only on the
system's path in parameter space, specifically the flux of some gauge field enclosed by that path.
%The field in question is a gauge field, akin to those found in quantum field theories, which
%arises naturally in Berry's formalism. 

Berry's work has since been applied to a diversity of phenomena, which can be broadly grouped under the umbrella of {\it geometric phases} or {\it topological phases} \cite{Shapere89,Anandan92}. Specific
instances of these geometric phases include the following: various analogues of the Aharonov-Bohm
effect \cite{Aharonov84,Casella90,He01}; the
rotation of the polarisation of light in twisted optical fibres, which was recognised by Pancharatnam
well before Berry's paper \cite{Pancharatnam56}; the so-called {\it molecular Aharonov-Bohm effect}
which introduces a gauge field to nuclear degrees of freedom in the Born-Oppenheimer
approximation to molecular dynamics \cite{Mead79,Mead80,Jackiw88,Mead92}; and even the dynamics of classical systems
such as low Reynolds number hydrodynamics \cite{Shapere89a}.

Geometric phases have also proved a fruitful avenue of investigation for mathematical
physicists due to their rich topological properties and their close connection with gauge theories
of quantum fields. Examples include mathematically formulating geometric phases
in terms of the holonomy of line bundles \cite{Simon83} and directly using the geometric phase to help
explain fractional statistics in the quantum Hall effect \cite{Arovas84} and the origin of Wess-Zumino
terms in theories of quantum chromodynamics \cite{Stone86}.
 \begin{figure}
\centering
\includegraphics[width=8.5cm]{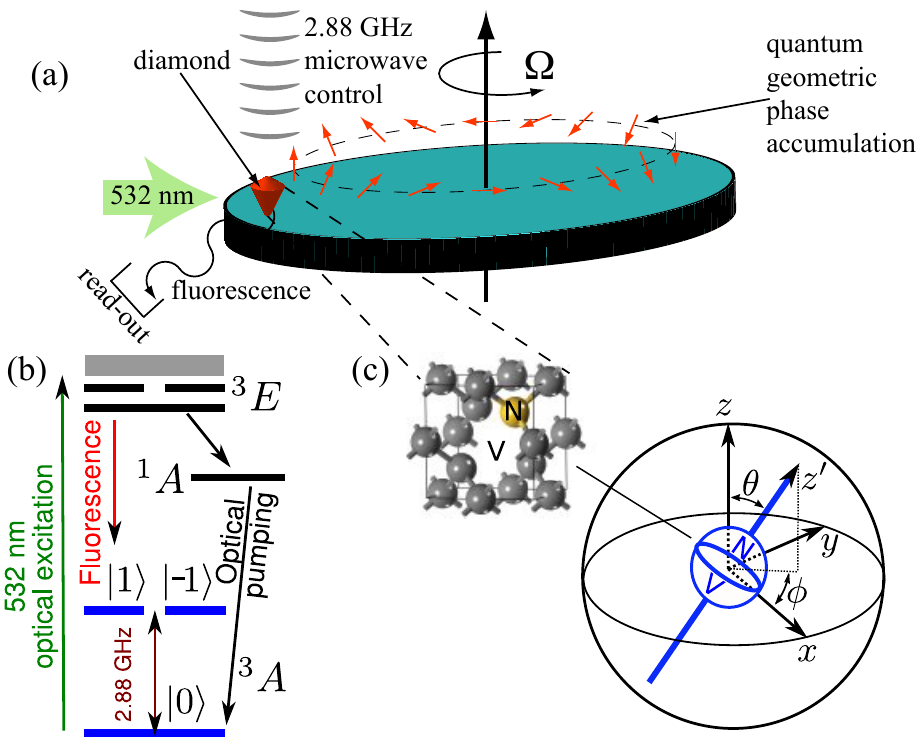}
\caption{(a) Schematic of system: a diamond crystal  containing NV centres, on a spinning table, with the microwave control polarized along the axis of rotation and the $532$ nm optical pulse used for readout. As the NV centre rotates around the axis the $m=\pm 1$ states can accumulate a geometric phase. An example of such phase accumulation is schematically shown for the $m=1$ state with the NV axis in the plane of rotation, i.e. $\theta=\pi/2$.  (b) Energy level diagram of the NV center. (c) Geometry  of the NV center. Defining the magnetic field direction of the microwave pulses as the $z$-direction, $z^{\prime}$ is the instantaneous direction of the NV axis, defined with respect to the lab-frame, unprimed coordinate system, by $\theta$ and $\phi$.  \label{zero}}
\vspace{-0.5cm}
\end{figure}
Despite this wealth of applications and observations, to date, only a few experimental observations of geometric phases due to mechanical rotation have been made \cite{Tycko87,Appelt94}. These measurements have been on ensembles of $^{35}$Cl \cite{Tycko87} and $^{131}$Xe \cite{Appelt94} nuclear spins. However,  no such measurement has been performed on an {\it individual} quantum system.   

Here we show how this might now be achieved using the quantum properties of a diamond, rotating around an axis, see Fig. 1(a). Such a measurement would represent a significant contribution to the foundations of quantum mechanics. The diamond nitrogen-vacancy (NV) system presents itself as an excellent tool for studying geometric
phases. The electron spin is the canonical quantum system and the NV center offers a
system in which a single spin can be initialised, coherently controlled, and measured.
It is also possible to mechanically move the diamond crystal, and the NV with it, about
some cyclical macroscopic trajectory within the spin coherence lifetime. As such this is an ideal system to investigate the accumulation of geometric phase, due to macroscopic rotation, on an {\it individual} quantum state.

The NV defect has a spin triplet ground state with a $2.88$ GHz zero-field splitting between the $m=0$ state and the degenerate $m=\pm 1$ inert states [Fig. 1(b)]. Optical excitation with $532$ nm light can pump the defect into the $m=0$ state and allows the population of the ground state to be read, since the $m=0$ state produces more fluorescence than the $m=\pm 1$ states \cite{Manson06,Steiner10}. The effective Hamiltonian of the ground state, ignoring crystal asymmetries and hyperfine effects, is
\begin{equation} 
H = \frac{1}{\hbar}D S_{z'}^2 +\frac{g\mu_B}{\hbar} \bm B \cdot \bm S. \label{Hamiltonian} 
\end{equation}
The first term is the zero-field splitting of the NV system itself, where $D$ is the zero-field splitting strength. It is this term which makes the crystal's orientation crucial. It defines an intrinsic quantisation direction $z^{\prime}$ (we reserve $z$ for the lab-frame coordinate) which lies along the axis connecting the nitrogen atom to its adjacent vacancy [Fig. 1(c)]. The second term is the usual Zeeman splitting interaction with a magnetic field. The magnetic field is $\bm B$, and $g \approx 2$ and $\mu_B$ are the Land\'e $g$ factor and the Bohr magneton respectively.
%\section{Geometric Phase in Diamond}

%We define the instantaneous eigenstates of the NV center's zero-field splitting Hamiltonian, projected onto the $z$ ($z^{\prime}$) axis as $| m\rangle_z$ ($| m\rangle_{z^{\prime}}$), where the $|0\rangle_{z^{\prime}}$ is the state which is produced by optical pumping and measured by recording fluorescence. A general state $|\psi\rangle$ is represented in matrix notation, with respect to the (lab-frame) $|m\rangle_z$ basis as:
 %\begin{eqnarray}
%| \psi \rangle=
%\left(
%\begin{array}{c}
 % _z \langle  1 | \psi \rangle \\
 % _z \langle  0 | \psi \rangle   \\   
%_z\langle  -1 | \psi \rangle
%\end{array}
%\right)_z.
 %\end{eqnarray}
 
 To evaluate the geometric phase for the NV center the instantaneous eigenstates of the Hamiltonian in terms of the adiabatically varied parameters $\theta$ and $\phi$ [see Fig. 1(b)] need to be determined. Writing the zero field splitting Hamiltonian as
 \begin{eqnarray}
 & &H_0=\frac{1}{\hbar}D S_{z'}^2 = \nonumber \\ & &D\hbar\left(
\begin{array}{ccc}
  \cos^2\theta +\frac{\sin^2\theta}{2} &\frac{e^{-i\phi}\cos\theta\sin\theta}{\sqrt{2}}& \frac{e^{-2i\phi}\sin^2 \theta}{2} \\
\frac{e^{i\phi}\cos\theta\sin\theta}{\sqrt{2}} & \sin^2\theta & -\frac{e^{-i\phi}\cos\theta\sin\theta}{\sqrt{2}} \\
 \frac{e^{2i\phi}\sin^2 \theta}{2}   & - \frac{e^{i\phi}\cos\theta\sin\theta}{\sqrt{2}} &  \cos^2\theta +\frac{\sin^2\theta}{2}
\end{array}
\right)_z
\nonumber \\
 \end{eqnarray}
with eigenstates $ \psi_{z^{\prime}}^{(m)}$:
 \begin{eqnarray}
 \psi_{z^{\prime}}^{(1)}&=&
 \left(
\begin{array}{c}
  \cos^2\frac{\theta}{2} \\
  \frac{e^{i\phi}}{\sqrt{2}}\sin\theta   \\   
e^{2i\phi}\sin^2\frac{\theta}{2}
\end{array}
\right)_z, \,\,\,\,\, 
 \psi_{z^{\prime}}^{(0)}=
 \left(
\begin{array}{c}
  -\frac{e^{-i\phi}}{\sqrt{2}}\sin\theta \\
 \cos\theta   \\   
\frac{e^{i\phi}}{\sqrt{2}}\sin\theta
\end{array}
\right)_z, \nonumber \\
 \psi_{z^{\prime}}^{(-1)}&=&
 \left(
\begin{array}{c}
  e^{-2i\phi}\sin^2\frac{\theta}{2} \\
  -\frac{e^{-i\phi}}{\sqrt{2}}\sin\theta   \\   
  \cos^2\frac{\theta}{2}
\end{array}
\right)_z.
\label{Eigen_z_prime}
 \end{eqnarray}
 
 \begin{figure}
\centering
\includegraphics[width=8.5cm]{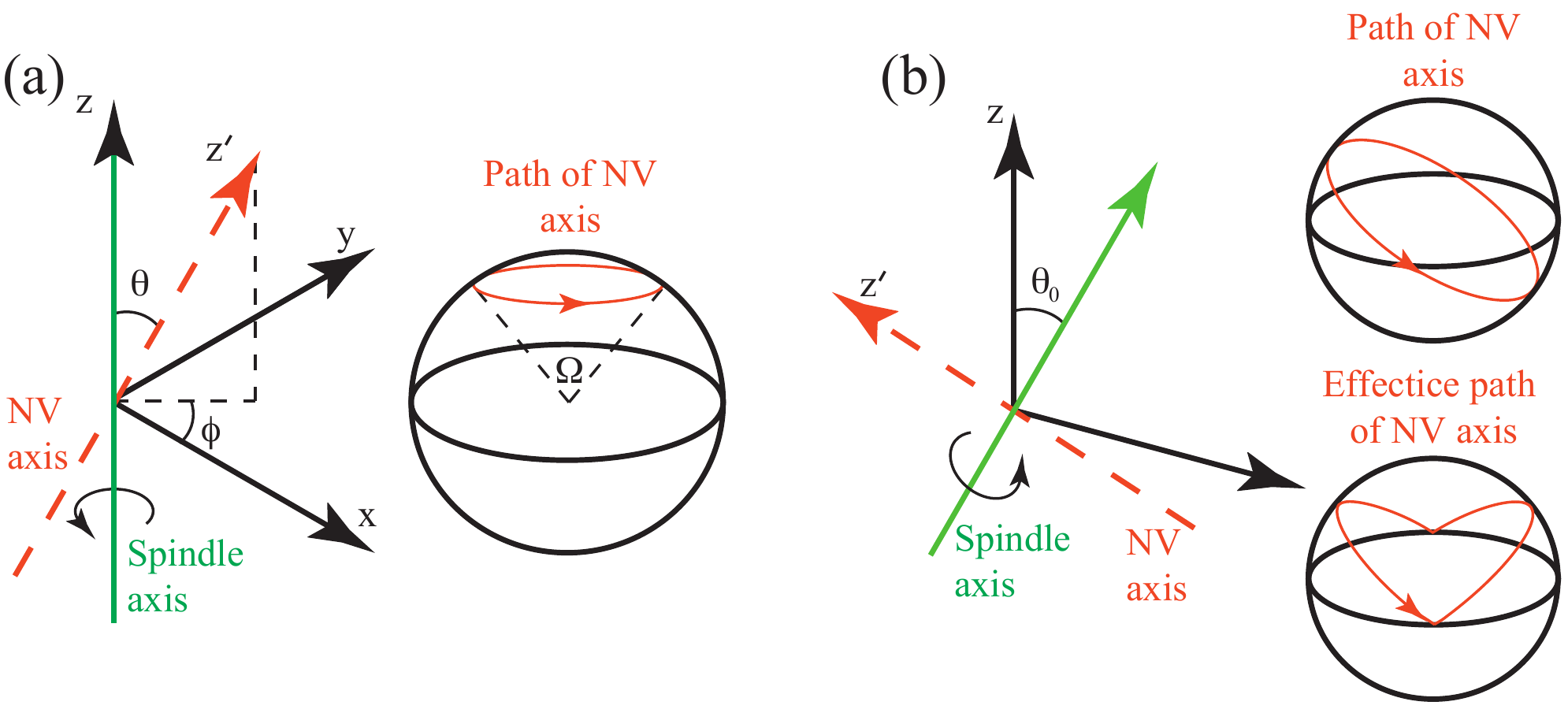}
\caption{(a) Proposed Ramsey  geometry. The crystal is attached
to a rotating spindle with the NV  ($z^{\prime}$) axis of one of its NV centers at an angle $\theta$ to the
spindle ($z$) axis. Rabi pulses are produced by a microwave field linearly polarised with
its magnetic field oscillating along the $z$-axis. The magnitude of the geometric phase after a
complete rotation of the spindle is given by the solid angle 
 subtended by the $z^{\prime}$-axis.
(b) Proposed spin echo geometry. The spindle is at an angle $\theta_0$ to the fixed $z$ axis
of the microwave field, and the NV axis is perpendicular to the spindle axis. The actual
path of the NV ($z^{\prime}$) axis encloses a solid angle of $2\pi$, which is not observable. The $\pi$-pulse
in the spin echo geometry, however, rectifies the alternating Berry phase accumulation,
producing a total phase of $4 \theta_0$ after a full spindle rotation, the solid angle enclosed by the
effective trajectory shown.  \label{one}}
\vspace{-0.5cm}
\end{figure}
 
The geometric phase, given initialisation into $ \psi_{z^{\prime}}^{(m)}$, after the crystal has been rotated along a trajectory $P$, defined as rotation about the $z$-axis with fixed $\theta$, is given by
 \begin{eqnarray}
 \Phi=\int_0^t \left[\frac{d \psi_{z^{\prime}}^{(m)}(t)}{dt} \right]^{\star} \psi_{z^{\prime}}^{(m)}(t) dt=\int_P m\left(1-\cos \theta\right)d\phi. \label{Berry}
\end{eqnarray}
A similar effect occurs if the quantization axis is determined by an electric field, which is rotated \cite{Book}. For a closed loop in parameter space, this is simply the solid angle enclosed by the trajectory of $z^{\prime}$, see Fig. 2(a). In the derivation of Eq.~(\ref{Berry}) we have assumed that the evolution of the system is adiabatic: $\dot{\theta}^2 +\sin^2\theta \dot{\phi}^2 \ll 2D^2$, i.e. the adiabatic approximation is valid provided the angular velocity of the crystal is much less that the zero-field splitting frequency, which is true for the cases we consider below.

While there is a gauge degree of freedom in defining the instantaneous eigenstates $\psi_{z^{\prime}}^{(m)}$, the geometric phase for a closed loop trajectory is independent of the choice of gauge. However, we are interested in a geometric phase for trajectories which are not closed loops and must then be careful in choosing our gauge.

The geometric phase is observed through interaction with a microwave magnetic field ${\bm B_R}$ and it is the phase difference between the NV center and the microwave field, as seen by the NV center, which is the observable quantity.  Consider a linearly polarized microwave field tuned to the zero-field splitting transition, oscillating along the fixed $z$ axis, ${\bm B_R}=B_R\cos(\omega t) \hat{z}$. The Hamiltonian for the interaction of this field with the NV center is
\begin{eqnarray}
 H_{{\rm int}}&=&\frac{g\mu_B}{\hbar}{\bm B_R} \cdot {\bm S} \nonumber \\
 &=& g \mu_B B_R \cos(\omega t)
 \left(
\begin{array}{ccc}
  \cos \theta  & \frac{e^{i\phi}\sin\theta}{\sqrt{2}} & 0 \\
\frac{e^{-i\phi}\sin\theta}{\sqrt{2}} & 0 & \frac{e^{i\phi}\sin\theta }{\sqrt{2}} \\
 0 &  \frac{e^{-i\phi}\sin\theta}{\sqrt{2}} &  -\cos\theta
\end{array}
\right)_{z^{\prime}}
\nonumber \\
&\approx& g \mu_B B_R \cos(\omega t)\frac{\sin\theta}{\sqrt{2}}
 \left(
\begin{array}{ccc}
  0 & e^{i\phi} & 0 \\
e^{-i\phi} & 0 & e^{i\phi} \\
 0 & e^{-i\phi}&  0
\end{array}
\right)_{z^{\prime}},
\label{H_int}
 \end{eqnarray}
 where the matrices are expressed with respect to the $|m\rangle_{z^{\prime}}$ basis as defined by Eq.~(\ref{Eigen_z_prime}). The approximation neglects the term proportional to $S_{z^{\prime}}$ and is valid for weak microwave fields: $g \mu_B B_R/\hbar \ll \omega$.
 
 Equation (\ref{H_int}) depends on both the polar $\theta$ and azimuthal $\phi$ angles. The dependence on the polar angle is simply a matter of effective strength, i.e. the effective microwave field strength experienced by the NV center is just $B_R \sin \theta$. The dependence on the azimuthal angle is more interesting: it has the character of a phase. The effect of a microwave pulse on state $\psi_{z^{\prime}}^{(m)}$ when the NV axis is at some azimuthal angle $\phi=\phi_0$ is the same as the effect when $\phi=0$ and the eigenstate is modified by a phase factor $\exp(-im_{z^{\prime}}\phi_0)$. Since the linearly polarised microwave field can be decomposed into two counter-rotating fields, it should be no surprise that the effective phase of the NV center should be so closely linked to the microwave field's angle relative the the NV axis.
 
This phase-like dependence of the interaction Hamiltonian on the NV orientation can be eliminated  via a gauge transformation, of the  $\psi_{z^{\prime}}^{(m)}$ basis, of the form
 \begin{eqnarray}
\psi_{z^{\prime}}^{(m)} \rightarrow e^{if_m(\theta,\phi)}\psi_{z^{\prime}}^{(m)}.
 \end{eqnarray}
Choosing $f_m =-m\phi$ eliminates the $\phi$-dependence of the interaction Hamiltonian [Eq.~(\ref{H_int})] as desired. This change of gauge does not alter measurable quantities. Using this gauge transformed eigenstate basis the geometric phase becomes
 \begin{eqnarray}
 \Phi=\int_P m\cos \theta d\phi. \label{Berry1}
 \end{eqnarray}  
The final state produced by the microwave pulse now depends only on the explicit phase given by Eq.~(\ref{Berry1}).

%\section{Experimental proposal}
The challenge is now to measure the relative phase accumulation between two states.
Because the bulk diamond NV electron spin has such long coherence times it should be possible
to observe a geometric phase by mechanically spinning a diamond crystal. The set-up we
propose is very similar to the Aharonov-Casher measurement proposed in Ref. \cite{Maclaurin09}. A diamond
crystal is mounted on a spinning spindle. Optical initialisation, coherent microwave
control, and fluorescence detection are used to measure the phase evolved after a certain
rotation angle. Below we consider two possible scenarios for measuring the geometric phase: %in Section A 
a relatively simple Ramsey geometry [Fig. 2(a)] and %in Section B 
a spin echo pulse sequence [Fig. 2(b)], enabling a significant increase in the maximum geometric phase observed at the cost of introducing additional $\pi$ microwave pulses. In each case it is assumed that the microwave pulse durations and readout times are of order $50$ ns and hence the diamond can be considered to be stationary.   %In Section C 
Finally the relative sensitivity of the measurements is investigated.

%\subsection{Proposed Ramsey Experimental Geometry} 
The Ramsey geometry is shown in Fig. 2(a), with the microwave field linearly
polarised with its magnetic field pointing along the spindle ($z$) axis. The crystal itself is mounted such that
the NV axis makes an angle $\theta$ to the spindle axis. To remove the degeneracy of the $\psi_{z^{\prime}}^{(\pm 1)}$ states we assume a magnetic
field rotating with the spindle, produced, for example, by a permanent magnet mounted
on the spindle.

For the Ramsey geometry as the
azimuthal angle $\phi$ of the NV axis passes zero, a $\pi/2$ Rabi pulse is applied, tuned to the
$\psi_{z^{\prime}}^{(0)} \rightarrow \psi_{z^{\prime}}^{(1)}$  transition. As the spindle continues to rotate, a relative geometric phase evolves between the $\psi_{z^{\prime}}^{(0)}$ and $\psi_{z^{\prime}}^{(1)}$ states, given by $\Phi=\phi\cos \theta$. Since the magnetic field, used to split the $\psi_{z^{\prime}}^{(\pm 1)}$ states, is static the only phase accumulated between the states is $\Phi$.
After the spindle has rotated through some angle $\phi_0$, a second $\pi/2$ pulse is applied, converting
the phase into a population difference, which is measured  by $532$ nm illumination and a fluorescence recording. A spindle rotation frequency of $\Omega = 4000\pi$ rad/s and a $10$ $\mu$s inhomogeneous broadening time ($T^*_2$) \cite{Jelezko04} would allow up to $20$ mrad of geometric phase to be accumulated. 

%\subsection{Proposed Spin Echo Experimental Geometry} 
To extend the coherence lifetime of the NV electron spin a spin echo pulse sequence could be employed, hence enabling a larger geometric phase to be measured. To do this a different geometry is required, shown in Fig. 2(b), in which the spindle axis is placed at an angle $\theta_0$ to the microwave field ($z$) axis, and the NV ($z^{\prime}$) axis is perpendicular to the spindle axis. The resulting geometric phase, alternates, and for a complete rotation is zero.

A spin echo control sequence, with $\pi$ pulses applied whenever the NV ($z^{\prime}$) axis is perpendicular to $z$, would rectify the alternating geometric phase, producing a total phase of $\Phi=4n\theta$, where $n$ is the number of complete spindle rotations. %This accumulated geometric phase difference between the  $\psi_{z^{\prime}}^{(0)}$ and $\psi_{z^{\prime}}^{(1)}$ states is, as in the Ramsey geometry, in addition to the phase difference accumulated due to the presence of external magnetic fields, Eq.~(\ref{magnetic_phase}), used to lift the degeneracy between the $m=\pm 1$ states. 
By extending the coherence time from $T_2^*$ to $T_2 \approx 2$ ms \cite{Balasubramanian09} a spin echo control sequence enables a geometric phase difference of  $4$ radians to be produced.

%\subsection{Relative Measurement Sensitivity} 
The measurements proposed above would consist of many repetitions of the pulse sequence
 to get an average signal. The sensitivity of the measurement of the geometric
phase is determined by the Poissonian statistics of spin projection, photon emission, and
photon collection. The uncertainty $\Delta \Phi$ in the measurement of  a geometric phase is related
to the uncertainty $\Delta S$ of the normalised fluorescence signal $S$,
\begin{eqnarray}
\Delta \Phi=\Delta S \left(\frac{dS}{d\Phi}\right)^{-1}=2\Delta S.
\end{eqnarray}
The second equality arises because the normalised signal, a sinusoidal function of $\Phi$, has a maximum gradient of $1/2$. By appropriately retarding the phase of the final $\pi/2$
pulse it can be ensured that the sinusoid is at its steepest point at the time of measurement.

The normalised signal $S$ is the number of photons collected over $N_r$ runs, normalised
so that $\langle S \rangle = 1/2$ when the populations of $\psi_{z^{\prime}}^{(0)}$ and $\psi_{z^{\prime}}^{(1)}$ are equal. If each measurement
of  $\psi_{z^{\prime}}^{(0)}$ or  $\psi_{z^{\prime}}^{(1)}$ corresponded to the emission and detection of exactly one or zero photons
respectively, then the variance of $S$ would be $(\Delta S)^2 = 1/(2N_r)$. A more careful analysis,
taking into account the statistics of (spontaneous) photon emission, imperfect detection 
and the nonzero fluorescence of the $\psi_{z^{\prime}}^{(1)}$ state, modifies the variance of the normalised
signal by a factor $C^2$, giving $(\Delta S)^2 = 1/(2C^2N_r)$. The physical basis for the factor $C$ ($C \approx 0.15$ for typical experiments and $C=1$ in
the ideal case) is described by Taylor {\it et al.} \cite{Taylor08}. 

The relative sensitivity for a series of geometric phase measurements, using a single NV center, is
\begin{eqnarray}
\frac{\Delta \Phi}{\Phi}\approx \frac{2\pi \sqrt{2T_M}}{C \Omega T_2^{(*)}\sqrt{T_T}},
\end{eqnarray}
where $T_M$ is the time to take a single measurement, $T_T$ is the total averaging time. Expressing $T_M$ in terms of the relevant de-coherence time ($T_M=aT_2^{(*)}$, where $a>1$) the relative sensitivity becomes
\begin{eqnarray}
\frac{\Delta \Phi}{\Phi}\sqrt{T_T}\approx \frac{2\pi \sqrt{2 a}}{C\Omega\sqrt{T_2^{(*)}}}.
\end{eqnarray}
Based on the proposed experimental parameters given above ($a=2$), this corresponds to a relative uncertainty in the measurement  of $\Phi$ with (without) spin echo pulses sequences of $0.15$ Hz$^{-1/2}$ ($2$ Hz$^{-1/2}$), or a $0.15$\% ($2$\%) uncertainty after three hours.

%\section{Conclusions}
An alternative approach to measuring a geometric phase using an NV centre is to make use of
an ancillary nuclear spin. Coherent control of nearby nuclear spins such as $^{13}$C or $^{14}$N has
been demonstrated experimentally \cite{Dutt07,Jiang09,Jelezko04a}. Hyperfine coupling between the electron and nuclear spin allows controlled-not (CNOT) operations to be performed, conditionally
flipping one spin depending on the value of the other spin, which allows information to be
exchanged between electron and nuclear spins. The adiabatically varying Hamiltonian in a nuclear spin geometric phase experiment could come from a number of sources. It is known that the $^{14}$N spin of an NV centre has a $5$ MHz zero-field nuclear quadrupole splitting along the NV axis, in complete analogy with the zero-field splitting of the NV electronic spin. The zero-field splitting of a $^{13}$C
spin, however, will depend on its location. One approach would be to split the nuclear spin states using a magnet mounted on the spindle which rotates with the crystal. The splitting of nuclear spin states will in any case be on the order of MHz, so adiabaticity (with kHz spindle rotation speed) is still maintained. Due to their weaker interaction (by three orders of magnitude) with the environment,
nuclear spins have longer coherence times than electronic spins, $T^*_2$ on the order of ms \cite{Dutt07} for $^{13}$C spin. A geometric phase experiment using nuclear spin would thus be both more
sensitive and would allow a complete rotation of the spindle within the coherence time at the expense of a more complicated pulse sequence.

We have demonstrated that a geometric phase shift manifests between the internal magnetic states of a single nitrogen-vacancy defect, within a rotating diamond crystal. The measurement of such a geometric phase shift in a macroscopically rotating single atom-scale quantum object would provide a unique test of our fundamental understanding of quantum mechanics. As such we have demonstrated that the measurement of geometric phase shifts of $>$ 1 radian in such systems is possible. The analysis presented above is not only important in terms of demonstrating geometric phase shifts in macroscopically rotating quantum systems it also provides the basis for quantifying geometric phase shift effects in the use of nano-diamonds as high precision translational and rotational sensors \cite{Dougal_Thesis,Mcquinness11,Dima,Paola}.

%\section{Acknowledgements} 
This research was supported by the Australian Research Council Centre of Excellence for Quantum Computation and Communication Technology (CE110001027). L.C.L.H. was supported under an Australian Research Council Professorial Fellowship (DP0770715).

%\begin{figure}
%\centering
%\includegraphics[width=8.5cm]{Fig_1_Andy}
%\caption{(a) Important timescales as a function crystal radius (assuming the crystal is immersed in room  temperature water): characteristic rotation time, $1/k_D$ (red dashed curve); population mixing time, $t_M$ (blue dotted curve); angular velocity damping time $t_d$ (green dashed-dotted curve).  (b) Measurement sensitivity assuming $C = 0.01$ and $10^{16}$cm$^{-3}$ NV density: DC magnetic field sensitivity (solid black  curve, right axis); rotation rate relative sensitivity from Berry's phase decoherence  (red dashed curve, left axis) and from non-adiabatic population mixing (dotted blue curve, left axis). \label{time_scales}}
%\vspace{-0.5cm}
%\end{figure}


\begin{thebibliography}{99}
\bibitem{Aharonov59} Y. Aharonov and D. Bohm, Phys. Rev. {\bf 115}, 485 (1959). 
  %\bibitem{Mcquinness11} L.P. Mcguinness, {\it et al.}, Nature Nanotechnology {\bf 6}, 358 (2011).
%\bibitem{Chamber60} R.G. Chambers, Phys. Rev. Lett. {\bf 5}, 3 (1960).
\bibitem{Berry84} M.V. Berry, Proc. R. Soc. London A {\bf 392}, 45 (1984). 
\bibitem{Shapere89} {\it Geometric Phases in Physics}, edited by A. Shapere and F. Wilczek (World Scientific, Singapore, 1989).
\bibitem{Anandan92} J. Anandan, Nature {\bf 360}, 307 (1992). 
\bibitem{Aharonov84} Y. Aharonov and A. Casher, Phys. Rev. Lett. {\bf 53}, 319 (1984).
\bibitem{Casella90} R.C. Casella, Phys. Rev. Lett. {\bf 65}, 2217 (1990). 
\bibitem{He01} X.-G. He and B.H.J. McKellar, Phys. Rev. A {\bf 64}, 022102 (2001).
\bibitem{Pancharatnam56} S. Pancharatnam, Proc. Ind. Acad. Sci. {\bf 44}, 247 (1956).
\bibitem{Mead79} C.A. Mead and D.G. Truhlar, J. Chem. Phys. {\bf 70}, 2284 (1979).
\bibitem{Mead80}  C.A. Mead, Chem. Phys. {\bf 49}, 23 (1980).
\bibitem{Jackiw88} R. Jackiw, Int. J. Mod. Phys. A {\bf 3} 285 (1988).
\bibitem{Mead92}  C.A. Mead, Rev. Mod. Phys. {\bf 64}, 51 (1992).
\bibitem{Shapere89a} A. Shapere and F. Wilczek, J. Fluid Mech. {\bf 198}, 557 (1989).
 \bibitem{Simon83} B. Simon, Phys. Rev. Lett. {\bf 51}, 2167 (1983). 
 \bibitem{Arovas84} D. Arovas, J.R. Schrieffer and F. Wilczek, Phys. Rev. Lett. {\bf 53}, 722 (1984).
 \bibitem{Stone86} M. Stone, Phys. Rev. D {\bf 33}, 1191 (1986).
 \bibitem{Tycko87} R. Tycko, Phys. Rev. Lett. {\bf 58}, 2281 (1987).
 \bibitem{Appelt94} S. Appelt, G. W\"{a}ckerle and M. Mehring, Phys. Rev. Lett. {\bf 72}, 3921(1994).
 \bibitem{Manson06} N.B. Manson, J.P. Harrison and M.J. Sellars, Phys. Rev. B {\bf 74}, 104303 (2006).
 \bibitem{Steiner10} M. Steiner, P. Neumann, J. Beck, F. Jelezko and J. Wrachtrup, Phys. Rev. B {\bf 81}, 035205 (2010).
 \bibitem{Book} D. Budker, D.F. Kimball and D.P. DeMille, {\it Atomic Physics: An Exploration Through Problems and Solutions}, $2^{nd}$ {\it Ed.} (Oxford University Press, New York, 2008).
 \bibitem{Maclaurin09} D. Maclaurin, A.D. Greentree, J.H. Cole, L.C.L. Hollenberg and A.M. Martin, Phys. Rev. A {\bf 80}, 040104(R) (2009).
  \bibitem{Jelezko04} F. Jelezko, T. Gaebel, I. Popa, A. Gruber and J. Wrachtrup, Phys. Rev. Lett. {\bf 92}, 076401 (2004).
  \bibitem{Balasubramanian09} G. Balasubramanian {\it et al.}, Nature Materials {\bf 383} (2009).
   \bibitem{Taylor08} J.M. Taylor {\it et al.}, Nature Physics {\bf 4}, 810 (2008).
  \bibitem{Dutt07} M.V.G. Dutt {\it et al.}, Science {\bf 316}, 1312 (2007).
  \bibitem{Jiang09}  L. Jiang {\it et al.}, Science {\bf 326}, 267 (2009).
  \bibitem{Jelezko04a} F. Jelezko {\it et al.}, Phys. Rev. Lett. {\bf 93}, 130501 (2004). 
  \bibitem{Dougal_Thesis} D. Maclaurin, {\it From Geometric Phases to Intracellular Sensing: New Applications of the Daimond Nitrogen-Vacancy Centre} (MPhil Thesis, University of Melbourne, 2010).
   \bibitem{Mcquinness11} L.P. Mcguinness, {\it et al.}, Nature Nanotechnology {\bf 6}, 358 (2011).
   \bibitem{Dima}M. Ledbetter, K. Jensen, R. Fischer, A. Jarmola and D. Budker, arXiv:1205.0093.
   \bibitem{Paola} A. Ajoy and P. Cappellaro, arXiv:1205.1494

 

\end{thebibliography}
\end{document}